\documentclass[aps,pra,twocolumn,showpacs,showkeys,preprintnumbers,amsmath,amssymb]{revtex4}

\usepackage{graphicx}
\usepackage{dcolumn}
\usepackage{bm}

\DeclareMathOperator*{\im}{Im}

\begin{document}


\title{Propagation of relativistic charged particles in ultracold atomic gases
\\with Bose-Einstein condensates}

\author{Yurii Slyusarenko}
\affiliation{Akhiezer Institute for Theoretical Physics, NSC KIPT, 61108 Kharkiv, Ukraine\\
Department of Physics and Technology, Karazin National University, 61077 Kharkiv, Ukraine}
\author{Andrii Sotnikov}
\email{sotnikov@th.physik.uni-frankfurt.de}
\affiliation{Akhiezer Institute for Theoretical Physics, NSC KIPT, 61108 Kharkiv, Ukraine\\
Department of Physics and Technology, Karazin National University, 61077 Kharkiv, Ukraine}
\affiliation{Institut f\"ur Theoretische Physik, Johann Wolfgang Goethe-Universit\"at, 60438 Frankfurt am Main, Germany}

\date{\today}

\begin{abstract}
We study theoretically some effects produced by a propagation of the charged particles in dilute gases of alkali-metal atoms in the state with Bose-Einstein condensates. The energy change of the high-speed (relativistic) particle that corresponds to the Cherenkov effect in the condensate is investigated. We show that in the studied cases the particle can both loose and receive the energy from a gas. We find the necessary conditions for the particle acceleration in the multi-component condensate. It is shown that the Cherenkov effect in Bose-Einstein condensates can be used also for defining the spectral characteristics of atoms.
\end{abstract}


\maketitle

\section{Introduction}

Physical studies devoted to the charged particles propagating in the refractive medium have a rather long history. To the most vivid effects in this case can be referred the Cherenkov radiation \cite{Che1934}, the ionization losses of energy \cite{Fer1940PR}, the nonlinear Doppler effect in the refractive medium \cite{Fra1942}, etc.
At the end of the past century experimentalists achieved the Bose-Einstein condensation (BEC) phase that is often named as a new state of matter. This state most commonly is observed in dilute gases consisting of alkali-metal atoms. In these systems the critical transition is achieved at ultralow temperatures (order of nanokelvin) and relatively low densities, $\nu\lesssim10^{14}$~cm$^{-3}$. Besides that, a Bose-Einstein condensate is known as a coherent state of matter, i.e., the state with practically similar behavior of atoms. Due to unusual requirements and properties, gases in the BEC state manifest unique effects that are inherent only to these systems (see Ref.~\cite{Pitaevskii2003} for details). The mentioned extreme conditions for the temperatures can result in more vivid manifestation of the physical phenomena that are well-studied for other systems. For example, to the one of these effects can be referred the ultraslow-light phenomenon in BECs \cite{Hau2004PRA,Sly2008PRA}. Therefore, now Bose-Einstein condensates are considered by many physicists as unique systems from the standpoint of the possibilities of the experimental observations and theoretical studies.

Taking into account all the mentioned facts, we set the problem of a propagation of the charged particles in a gas with Bose-Einstein condensates. Within this paper, we focus on some aspects of this problem.

\section{Formalism}

It is known that alkali-metal atoms are most effectively used for the studies of the phase with a BEC in atomic gases. For a description of the electromagnetic properties of dilute gases in some cases one may use the ideal gas approximation (see Ref. \cite{Sly2008PRA} for details). As it is easy to see, this model is also convenient for studying the effects that take place during the interaction of the charged particle with a BEC in vapors of alkali-metal atoms. For gases with comparably low densities of atoms, $\nu\lesssim10^{14}$~cm$^{-3}$, the microscopic approach is developed \cite{Sly2008PRA}. In the framework of the mentioned model it is shown that the dispersion characteristics of a gas with a good accuracy can be described by analysing the relation for its permittivity. In the case of the system considered in the BEC state, the Fourier transform of the permittivity can be written as follows:
\begin{equation}\label{eq.a12-1}
    \epsilon^{-1}(\textbf{k},\omega)=1+\dfrac{4\pi}{k^2}\sum\limits_{a,b}
    \dfrac{(\nu_a-\nu_b)|\sigma_{ab}(\textbf{k})|^2}
    {\omega-\Delta\varepsilon_{ab}+i\Gamma_{ab}/2},
\end{equation}
where $\omega$ and $\textbf{k}$ are the wave frequency and wave vector, respectively (here and below we set $\hbar=1$), indices $a$ and $b$ denote the sets of quantum numbers that correspond to the certain states of alkali-metal atoms. The quantities $\nu_a$ and $\nu_b$ are the atomic densities in these states, $\Delta\varepsilon_{ab}$ is the transition energy between defined states, and $\Gamma_{ab}$ is the natural linewidth that corresponds to the probability of the spontaneous transition between the states. The quantity $\sigma_{ab}(\textbf{k})$ in Eq.~(\ref{eq.a12-1}) is the matrix element of the charge density of atoms that is defined in the framework of the approximate formulation of the second quantization method in the case of the presence of the bound states (atoms) in the system (see also Ref. \cite{Pel2005JMP} for details),
\begin{eqnarray}\label{eq.a12-1.2}
    \sigma_{ab}(\textbf{k})
    =e\int d\textbf{y}\varphi_{a}^{*}(\textbf{y})
    \varphi_{b}(\textbf{y})
    \left[\exp{\left(i\frac{m_{p}}{m}
    \textbf{k}\textbf{y}\right)}\right.\nonumber
    \\
    \left.-\exp{\left(-i\frac{m_{e}}{m}
    \textbf{k}\textbf{y}\right)}\right].
\end{eqnarray}
Here $\varphi_{a}$ is the wave function of an atom in the quantum state $a$, $m_{p}$ and $m_{e}$ are the masses of the atomic core and electron, respectively ($m=m_{p}+m_{e}$).

A quantity of the energy that is absorbed by the gas in the interval of wave frequencies $d\omega$ and wave numbers $d\textbf{k}$ during a propagation of the charged particle can be described by the formula (see, e.g., Ref~\cite{Akhiezer1981}):
\begin{equation}\label{eq.a12-2}
    Q_{\omega\textbf{k}}=-\dfrac{2}{(2\pi)^4}\im
    \left(\dfrac{4\pi}{\omega\epsilon}|\textbf{j}_\parallel|^2
    +\dfrac{4\pi\omega|\textbf{j}_\perp|^2}{\omega^2\epsilon-c^2k^2}
    \right),
\end{equation}
where $\textbf{j}_\parallel$ and $\textbf{j}_\perp$ are the longitudinal and transversal components of the density of the external current ($\textbf{j}=\textbf{j}_\parallel+\textbf{j}_\perp$, $\textbf{j}_\parallel=(\textbf{kj})\textbf{k}/k^2$). Note that here and below for the sake of simplicity in the derived relations we set the magnetic permeability of the gas equal to unity.

In the case of the smallness of the energy losses (below we make estimates that confirm this approximation), the particle motion can be considered as uniform. Thus the current density $\textbf{j}$ that is introduced in Eq.~(\ref{eq.a12-2}) can be defined as:
\begin{equation*}
    \textbf{j}(\textbf{x},t)=e\textbf{v}\delta(\textbf{x}-\textbf{v}t),
\end{equation*}
where $e$ and $\textbf{v}$ are the particle charge and velocity, respectively, $\delta(x)$ is the Dirac delta-function. Therefore, the Fourier transform of the current density produced by the propagating particle has the form:
\begin{equation}\label{eq.a12-3}
    \textbf{j}(\omega,\textbf{k})=2\pi e\textbf{v}\delta(\omega-\textbf{kv}).
\end{equation}

Then, we can use Eq.~(\ref{eq.a12-3}) for finding the quantity (\ref{eq.a12-2}). After simple mathematical transformations one gets the relation that defines the energy change of the charged particle per unit of time, $\mathcal{E}_{\omega\textbf{k}}=-Q_{\omega\textbf{k}}/T$,
\begin{equation}\label{eq.a12-5}
    \mathcal{E}_{\omega\textbf{k}}=\dfrac{e^2\omega}{\pi^2}\delta(\omega-\textbf{kv})
    \im\dfrac{v^2/c^2-1/\epsilon}{\omega^2\epsilon/c^2-k^2}.
\end{equation}
To get this result, it is necessary to use also the formula:
\begin{equation*}\label{eq.a12-4}
    \delta^2(\omega-\textbf{kv})=\dfrac{T}{2\pi}\delta(\omega-\textbf{kv}),
\end{equation*}
where $T$ is the flight time of the particle through the system under consideration.

Hence, the total change of the energy per unit of length is given by
\begin{equation}\label{eq.a12-6}
    \dfrac{d\mathcal{E}}{dx}=\dfrac{1}{v}\int d\omega d^3k\mathcal{E}_{\omega\textbf{k}}.
\end{equation}
It should be noted that, in general case, the charged particle can both loose the energy (the case of slowing, $d\mathcal{E}/dx<0$) and receive the energy
(the case of acceleration, $d\mathcal{E}/dx>0$). As one can see from Eq.~(\ref{eq.a12-5}), the realization of the mentioned cases depends on the initial velocity of the particle and dispersion characteristics of the system. In the next section we study in detail some possibilities of the realization of these cases in ultracold gases of alkali-metal atoms with Bose-Einstein condensates.

\section{Energy change of the charged particle that propagates in a BEC}

Note that dilute gases of alkali-metal atoms have an index of refraction close to unity even in the resonance regions. In particular, the change of the refractive index of a gas of sodium atoms with the density $\nu\sim10^{13}$ cm$^{-3}$ in the frequency region corresponding to the $D_2$ line is relatively small, $\Delta n \sim0.01$. Therefore, the effects resulting from the pole $1/\epsilon$ (see Eq. (\ref{eq.a12-5})) do not appear in this system. Evidently, in some particular cases, another pole, $\omega^2\epsilon-c^2k^2=0$, can play a key role. It should be noted that this pole corresponds to a contribution of the Cherenkov effect in the energy change of the charged particle.

We must emphasize that the Cherenkov radiation in the dispersive medium is observed only in the region of the frequencies with the following condition:
\begin{equation}\label{eq.a12-7}
    n(\omega)\beta\geq1,
\end{equation}
where $\beta=v/c$,
\begin{equation}\label{eq.a12-7.2}
    n^2(\omega)=(\sqrt{\epsilon'^2+\epsilon''^2}+\epsilon')/2,
\end{equation}
$\epsilon'$ and $\epsilon''$ are the real and imaginary parts of the permittivity, respectively, $\epsilon=\epsilon'+i\epsilon''$.

Hence, we conclude that in a medium with $n(\omega)\approx1$ the energy (\ref{eq.a12-5}) changes only at the relativistic values of the velocity of the propagating particles. As it is easy to see from direct calculations, e.g., for the maximal value $n_{\max}=1.01$ it is necessary to accelerate particles up to velocities $v\approx0.99c$ that correspond to the Lorentz factor $\gamma\approx7$. In other words, the Cherenkov effect in the system under consideration can be observed only for the particles with the kinetic energy that, at least, is greater than the rest energy in several times.

To analyse the peculiarities of the energy change of the charged particle in an atomic BEC, we use the model of a two-level system. To this end, we fix in Eq.~(\ref{eq.a12-1}) the indices corresponding to the chosen quantum states in atoms, $a=1$ and $b=2$ ($\varepsilon_1<\varepsilon_2<0$). In other words, we consider the case when the frequency of the potential radiation is close to the energy spacing between the defined states. As a result, from Eq.~(\ref{eq.a12-1}) one gets
\begin{equation}\label{eq.a12-8}
    \epsilon(\textbf{k},\omega)=\left[1+\dfrac{4\pi}{k^2}
    |\sigma(\textbf{k})|^2\dfrac{(\nu_1-\nu_2)}
    {\delta\omega+i\Gamma/2}\right]^{-1},
\end{equation}
where $\delta\omega$ is the frequency detuning,
$\delta\omega=\omega-\varepsilon_1+\varepsilon_2$. Note that in the quantities $\sigma(\textbf{k})$, $\delta\omega$ and $\Gamma$ we omit the pair index \textquotedblleft 12'' that corresponds to the chosen resonant transition in the atom. Then, the permittivity (\ref{eq.a12-8}) can be divided into the real and imaginary parts. Within the approximation $(\epsilon'-1)\ll1$, the energy change (\ref{eq.a12-5}) is written in the form:
\begin{equation}\label{eq.a12-9}
    \mathcal{E}_{\omega\textbf{k}}\approx-\epsilon''\dfrac{e^2\omega^3\beta^2}{\pi^2c^2}
    \left[\left(\dfrac{\omega^2\epsilon'}{c^2} -k^2\right)^2
    +\left(\dfrac{\omega^2\epsilon''}{c^2}\right)^2\right]^{-1},
\end{equation}
where
\begin{equation}\label{eq.a12-9.2}
    \epsilon'=\dfrac{\delta\omega(\delta\omega+\alpha)+\Gamma^2}
    {(\delta\omega+\alpha)^2+\Gamma^2},\quad
    \epsilon''=\dfrac{\alpha\Gamma}
    {(\delta\omega+\alpha)^2+\Gamma^2},
\end{equation}
\begin{equation}\label{eq.a12-9.3}
    \alpha(\textbf{k})=\dfrac{4\pi}{k^2}
    |\sigma(\textbf{k})|^2{(\nu_1-\nu_2)}.
\end{equation}

\begin{figure}
\includegraphics[width=.8\linewidth,keepaspectratio]{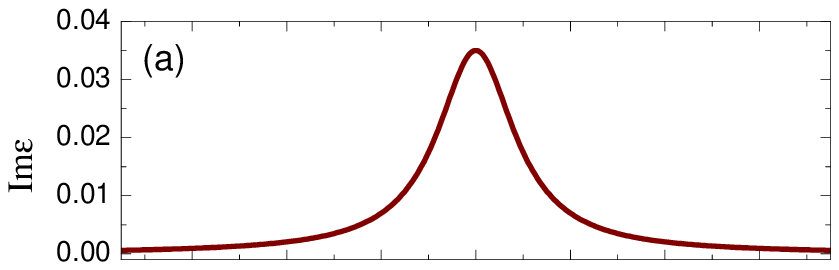}
\includegraphics[width=.18\linewidth,keepaspectratio]{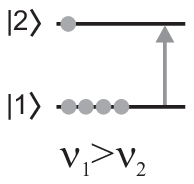}\\
\includegraphics[width=.8\linewidth,keepaspectratio]{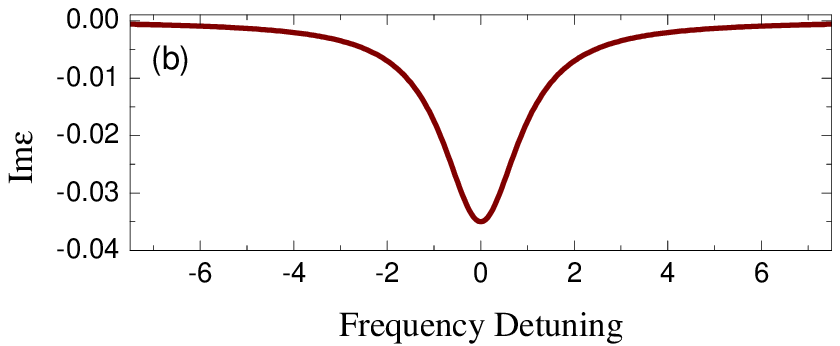}
\includegraphics[width=.18\linewidth,keepaspectratio]{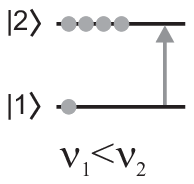}
  \caption{(Color online)
  Dependencies of the imaginary part of the permittivity on the frequency detuning ($\Delta=2\delta \omega/\Gamma$) in the case of the normal (a) and inverse (b) population of the two-level system.}
  \label{fig.01}
\end{figure}

The relation (\ref{eq.a12-9}) contains an important information about the energy change of the charged particle in the two-level system. It should be mentioned that the sign of the quantity  $\mathcal{E}_{\omega\textbf{k}}$ depends on the sign of the imaginary part $\epsilon''$ of the permittivity. This quantity, in turn, is proportional to the occupation difference $(\nu_1-\nu_2)$ of the quantum states in the system [see Eqs. (\ref{eq.a12-9.2}) and (\ref{eq.a12-9.3})]. Therefore, we come to a result that there are two particular cases for the energy change of the charged particle. In the system with the normal population, $\nu_1>\nu_2$, the Cherenkov losses can be observed, $\mathcal{E}_{\omega\textbf{k}}<0$. Otherwise, in the system with the inverse population, $\nu_1<\nu_2$, the particle can be accelerated by a condensate. The dependencies of the imaginary part of the permittivity on the frequency detuning for the both cases are shown in Fig.~\ref{fig.01}.

\begin{figure}
\includegraphics[%
  width=1.0\linewidth,
  keepaspectratio]{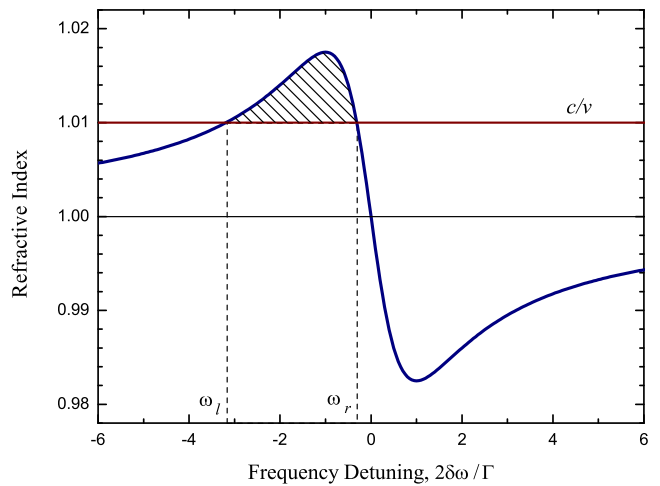} \caption{(Color online)
  Dependence of the refractive index on the frequency detuning.
  The hatched area corresponds to the region of the observation of the Cherenkov effect.}
  \label{fig.02}
\end{figure}

Then, we use Eqs.~(\ref{eq.a12-6}) and (\ref{eq.a12-9}) for defining the total energy change in the resonance region. To this end, the condition (\ref{eq.a12-7}) must be taken into account. This relation characterizes an appearance of the Cherenkov effect in the system. After the integration over the angles and absolute value of the wave number, we come to a result
\begin{equation}\label{eq.a12-10}
\begin{split}
    \dfrac{d\mathcal{E}}{dx}=-\dfrac{2 e^2}{\pi v^2}\int\limits_{\omega_l}^{\omega_r}d\omega\omega
    \epsilon''(\omega)\int\limits_{0}^{\theta_0(\omega)}
    \dfrac{\sin\theta d\theta}{\cos^3\theta}
    \\
    \times\left[\left(\epsilon'-\dfrac{c^2}{v^2\cos^2\theta}\right)^2
    +\epsilon''^2\right]^{-1},
\end{split}
\end{equation}
where the limiting angle $\theta_0(\omega)$ can be found from the condition
$n(\omega)\beta\cos\theta_0=1$, and the quantities $\omega_{l,r}$ define the limits of the frequency interval where the effect is observed, $n(\omega_{l,r})\beta=1$ (see also Fig.~\ref{fig.02}).

After integrating of the internal integral over polar angle one gets
\begin{equation}\label{eq.a12-11}
\begin{split}
    \dfrac{d\mathcal{E}}{dx}=-\dfrac{e^2}{\pi c^2}\dfrac{\epsilon''}{|\epsilon''|}
    \int\limits_{\omega_l}^{\omega_r}d\omega\omega
    \left\{\arctan\left[\dfrac{\beta^2(\epsilon'^2
    +\epsilon''^2)-\epsilon'}{|\epsilon''|}\right]\right.
    \\
    \left.-\arctan\left[\dfrac{\beta^2(\epsilon'^2
    +\epsilon''^2)-n^2\epsilon'}{|\epsilon''|n^2}\right]\right\}.
\end{split}
\end{equation}
Therefore, Eq.~(\ref{eq.a12-11}) describes the energy change of the charged particle on the frequency interval that corresponds to the resonant transition in the gas of atoms forming BEC. Naturally, to calculate the total energy change, it is necessary to sum the separate contributions of the resonant transitions where the condition (\ref{eq.a12-7}) is satisfied.

For the sake of simplicity and visibility we do not consider the particular cases with a large number of resonant transitions in alkali-metal atoms. To get the qualitative results, one can estimate the energy change by the order of magnitude in the region of a single resonance. Considering that the difference of the arctangents in the integrand of Eq.~(\ref{eq.a12-11}) does not exceed the $\pi$-number value and using the approximation $(\omega_r-\omega_l)\approx\Gamma$, $\Gamma\ll\omega_0$, where $\omega_0$ is the frequency corresponding to the energy of the resonant transition, $\Gamma\equiv\Gamma_{12}$, we find
\begin{equation}\label{eq.a12-12}
    \dfrac{d\mathcal{E}}{dx}\sim\dfrac{e^2}{c^2}\omega_0\Gamma.
\end{equation}
Basing on this approximate formula, it is easy to make estimates for the electron propagating in a gas of alkali-metal atoms. For example, for a gas of sodium atoms we can use the data that corresponds to the resonant $D_2$ line, $\Gamma\sim10^8$~s$^{-1}$ and $\omega_0\sim10^{15}$~s$^{-1}$. Thus, from Eq.~(\ref{eq.a12-12}) we find ${d\mathcal{E}}/{dx}\approx10^{-5}$ eV/cm. Next, basing on the fact that typical sizes of atomic samples with condensates are of the order of submillimeters, we finally get the change of energy corresponding to a single resonance (the transition $a\leftrightarrow b$) in an ultracold gas with a BEC, ${\Delta\mathcal{E}_{ab}}\approx10^{-7}$ eV.

It should be noted that, in more general case, the atoms must be considered as multilevel systems. Therefore, as it is mentioned above, the effect takes place for an ensemble of the resonant transitions where the relation (\ref{eq.a12-7}) is satisfied. But even an account of a relatively large (but, naturally, a finite) number of quantum states do not change significantly the kinetic energy $\mathcal{E}_{0}$ of a particle. Really, taking the Lorentz-factor value $\gamma=7$, one gets the ratio for the energies, $\Delta\mathcal{E}_{ab}/\mathcal{E}_{0}\sim10^{-11}$.

We also note that the effect can be enhanced by using more dense gases (this condition results in increasing of the hatched area in Fig.~\ref{fig.02}) and also by using multiple-ionized atoms or nuclei with the charge value $Ze$ that enhances the effect in $Z^2$ times.

\subsection{Possibility of the particle acceleration in a BEC}

It is shown above that the energy change of the propagating particle in the system under consideration is small. Probably, the changes in the velocity value resulting from the Cherenkov effect can not be directly measured for the case of dilute gases. But, in our opinion, an additional attention should be paid to a principal possibility of the particle acceleration by an ultracold atomic gas. It is mentioned above that the particle acceleration in the two-level model can be reached for the inverse-populated atomic states. As for the physical interpretation of this phenomenon, the analogy from the laser physics can be used. One can say that the propagating particle stimulates a transition from the upper to the lower state, and a part of the released energy is transferred to the charged particle resulting in its acceleration.

\begin{figure}
\includegraphics[%
  width=1.0\linewidth,
  keepaspectratio]{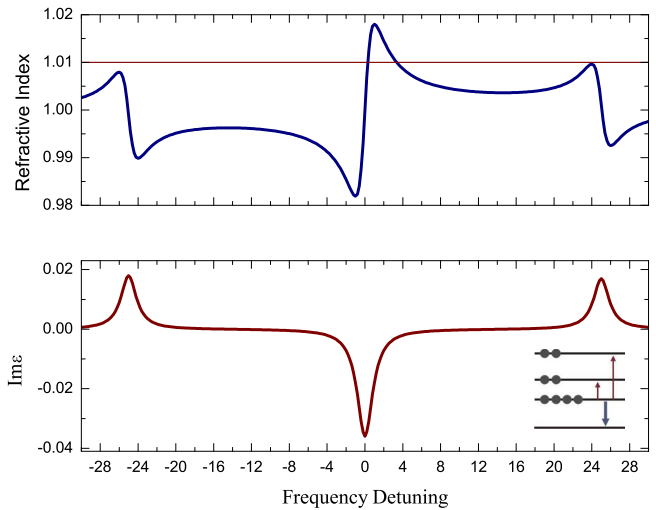} \caption{(Color online)
  Dispersion characteristics of a gas that accelerates the particle on the resonant transition with the inverse population. The density difference for this transition is taken twice larger than for the transitions with normal population (for sake of simplicity the linewidths
  $\Gamma_{ab}$ set the same).}
  \label{fig.03}
\end{figure}

Going beyond the two-level approximation, it is easy to see that on certain resonant transitions the particle can receive the energy from a medium and on other transitions it can emit the energy. Therefore, the particle interacts with a set of quantum transitions and looses the energy in general case. But in some particular cases, by a special tuning of the particle velocity and occupations of the atomic states, it is possible to accelerate the charged particle by the gas. This particular case is shown in Fig.~\ref{fig.03}. As one can see, in the region corresponding to the transition with the inverse population, the condition~(\ref{eq.a12-7}) is satisfied. At the same time, the velocity of the particle is not enough for the Cherenkov effect on the transitions with the normal population.

Apparently, the most appropriate quantum states for the experimental realization of this effect are the levels of the hyperfine structure of the ground state of alkali-metal atoms. This statement is based on the fact that these states have large lifetimes in comparison with the levels with the nonzero orbital moment. From the standpoint of the present experimental conditions, it is not difficult to populate the hyperfine states in the way to provide the particle acceleration (see the scheme in Fig.~\ref{fig.03}). If for the the upper (optical) transitions in this system the condition (\ref{eq.a12-7}) is not satisfied, the particle will be accelerated by a multi-component BEC. Otherwise, it receives the energy on the microwave transitions and emits the energy on the optical transitions.

Also we should note the case when the inverse population is provided by means of the optical pumping. This effect is usually produced by the additional lasers that are tuned up close to the resonant transitions. The effect in this case could be even larger according to the estimates in Eq.~(\ref{eq.a12-12}). But the dispersion characteristics of a gas in the case of optical pumping can not be studied in an appropriate way in the framework of the present paper. Really, the influence of additional strong fields result in the repopulation of the quantum states and, thus, it is necessary to go beyond the limits of the linear response theory \cite{Sly2008PRA}, which is used in this paper.

Therefore, it is shown that the particle that has a large kinetic energy (it can be larger than the rest energy in several times), in principal, can additionally receive the energy from ultracold gases consisting of atoms that have extremely small (or even zero) values of the kinetic energy.

\section{Defining the spectral characteristics of the atoms}

By registering of the Cherenkov radiation produced by the propagation of the charged particle in a BEC, it is possible to define the spectral characteristics of atoms that form this condensate. In particular, in this section, we use the Cherenkov effect for finding the natural linewidth $\Gamma$ value. This quantity can be defined both for the optical (dipole-allowed) and microwave (dipole-forbidden) transitions in alkali-metal atoms. The last problem may be actual also from the standpoint of the atomic-clock experiments (see, e.g., Ref.~\cite{Kas1989PRL}).

The main idea of this method consists in defining the frequency that corresponds to the maximum of the refractive index in the region of the certain resonant transition. This problem can be solved by varying the initial velocity of the propagating particle. As one can see from Fig. \ref{fig.02}, by reducing of the particle velocity, the height of the hatched area become smaller, and in the limit $1/\beta\rightarrow n_{\max}$ we get $\omega_{l,r}\rightarrow\omega_{c}$, $\omega_{c}=\omega(n_{\max})$.

Here and below we consider the case of the transmittance region, i.e., $|\epsilon''|\ll\epsilon'\approx1$. As one can see, in the zero order of the perturbation theory over $(\epsilon''/\epsilon')\ll1$, from Eq. (\ref{eq.a12-7.2}) it comes that $n(\omega)\approx\sqrt{\epsilon'(\omega)}$. Thus we get the relation $\omega_{c}^{(0)}=\omega_{0}-\Gamma/2$ from the condition for the maximum, $\partial n/\partial\omega=0$, and Eq.~(\ref{eq.a12-8}). The mentioned expression, in turn, transforms to the following formula:
\begin{equation}\label{eq.a12-15}
    \Gamma^{(0)}=2(\omega_{0}-\omega_{c}).
\end{equation}
Therefore, the linewidth $\Gamma^{(0)}$ can be found in the case when the frequency $\omega_{c}$ is defined from the spectral peak of the Cherenkov radiation in a BEC. Note that to get tangible results from Eq.~(\ref{eq.a12-15}), it is necessary also to use the $\omega_{0}$ value that must be determined with a relatively good accuracy.

In the case when the transition frequency $\omega_{0}$ is not known \emph{a priori}, the linewidth is possible to define by the use of an additional condition $n_{\max}=1/\beta$. Then, in accordance with Eq.~(\ref{eq.a12-9.2}), one gets
\begin{equation}\label{eq.a12-16}
    \Gamma^{(0)}=\dfrac{\alpha(\textbf{k})}{2(1/\beta-1)}.
\end{equation}
Hence, it is possible to find the natural linewidth $\Gamma^{(0)}$ from the microscopic characteristics of the system. As it is easy to see from Eqs. (\ref{eq.a12-1.2}) and (\ref{eq.a12-16}), in this case one need to know the explicit form for the wave function in two certain quantum states.

It should be noted that it is not difficult to find more explicit relations for the linewidth $\Gamma$ by the use of the terms appearing from the further expansion into series in the perturbation theory over $(\epsilon''/\epsilon')\ll1$. Note that this strict inequality corresponds to the approximation $|\alpha(k)|\ll\Gamma$. Thus, after some mathematical transformations, we come to the following expression:
\begin{equation}\label{eq.a12-17}
    \Gamma^{(1)}=2(\omega_{0}-\omega_{c})+3\alpha(\textbf{k})/2.
\end{equation}
Using the similar expansions for the undefined $\omega_{0}$ value, one gets
\begin{equation}\label{eq.a12-18}
    \Gamma^{(1)}=\dfrac{\alpha(\textbf{k})}{2(1/\beta-1)}
    \left[1-\dfrac{(1/\beta-1)}{2}
    \right].
\end{equation}

We must emphasize that basing on the Cherenkov effect in a BEC it is possible to define also other characteristics of the system under consideration. For example, using the described method  with the defined value of the natural linewidth one can find the transition frequency~$\omega_{0}$, the density~$\nu_a$ of atoms in the certain quantum states, or the velocity~$v$ of the propagating particle.

\section{Conclusion}
In this paper we investigate some effects that occur during the propagation of the charged particle in Bose-Einstein condensates of alkali-metal atoms. It is shown that it is necessary to use the charged particles with high (relativistic) velocities for appearing of the Cherenkov effect in the system. This condition corresponds to the fact that the refractive index for dilute atomic gases is close to unity even in the regions of resonant frequencies.

Basing on the relations for the permittivity of a gas in the BEC state, we find the energy change of the charged particle. It is shown that the effect of deceleration (acceleration) of the particle is very small for this system. Principal scheme for the preparation of the ultracold atomic gas that accelerates the relativistic charged particle is studied.

Basing on the Cherenkov effect we also consider the possibility for defining the spectral characteristics of atoms that form condensate. It is shown that the region of the local maximum of the refractive index can be found by varying the initial velocity of the charged particle.  We propose a method for finding the natural linewidth value from the Cherenkov radiation in a BEC basing on the knowledge of the additional spectral or microscopic characteristics of the system.

\acknowledgments{This work is partly supported by the National
Academy of Sciences of Ukraine, Grant No. 55/51-2010.}

\end{document}